\documentclass{aa}

\usepackage{txfonts}
\usepackage{graphicx}

\begin{document}

\title{Swift multi-wavelength observations of the bright flaring burst
GRB~051117A.}


\author{M. R. Goad\inst{1}, K. L. Page\inst{1}, O. Godet\inst{1},
  A. Beardmore\inst{1}, J.P. Osborne\inst{1}, P.T. O'Brien\inst{1},
  R. Starling\inst{1}, S. Holland\inst{2}, D. Band \inst{2}, 
  A. Falcone\inst{3}, N. Gehrels\inst{2},  D.N. Burrows\inst{3},
  J.A. Nousek\inst{3}, P.W.A. Roming\inst{3}, A. Moretti\inst{4}, and
  M. Perri\inst{5}.}

\institute{Department of Physics and Astronomy, University of Leicester,
LE1 7RH, UK
\and NASA Goddard Space Flight Center, Greenbelt, MD 20771, USA
\and Pennsylvania State University, University Park, PA 16802, USA
\and INAF-Osservatorio Astronomica di Brera, via Bianchi 46, 23807
\and ASI Science Data Center, Via Galileo Galilei, I-00044 Frascati,
Italy.
}

\date{Received : / Accepted : }

\abstract{We report on the temporal and spectral characteristics of the early
X-ray emission from the Gamma Ray Burst 051117A as observed by {\it
Swift\/}. The superb quality of the early X-ray light-curve and spectra of
this source, one of the brightest seen by the X-ray Telescope at such early
times, allows an unprecedented look at the spectral and temporal evolution of
the prompt and early afterglow emission for this GRB and allows us to place
stringent limits on the detection of lines.  GRB~051117A displays a highly
complex light-curve, with an apparent initial slow decline of slope
$\alpha=0.77\pm0.07$ ($f(t)\propto t^{-\alpha}$) dominated by numerous
superposed flares of varying amplitude and duration. Between orbits 2 and 3,
the X-ray light-curve drops abruptly, highlighting the dominance of flaring
activity at early times, and indicating that the central engine for this burst
remains active for several kiloseconds after the initial explosion. The late
time slope ($t>10^{4}$~s) also decays relatively slowly with a powerlaw index
of $\alpha=0.66$, breaking to a steeper slope of 1.1, 170~ks after the BAT
trigger.  The X-ray light-curve at early times is characteristic of a noise
process,  consisting of random shots superposed on an underlying powerlaw
decay, with individual shots well-modelled by a fast-rise and exponential
decay spanning a broad range in rise-times and decay rates. A temporal
spectral analysis of the early light-curve shows that the photon index and
source intensity are highly correlated with the spectrum being significantly
harder when brighter, consistent with the movement of the peak of the Band
function to lower energies following individual flares.  The high quality
spectrum obtained from the first orbit of WT mode data, enables us to place a
$3\sigma$ upper limit on the strength of any emission line features of EW$<
15$eV, assuming a narrow emission-line of 100~eV at the peak of the effective
area.

\keywords{gamma-ray: bursts -- Gamma-rays, X-rays: individual (GRB~051117A)}}

\titlerunning{Swift observations of GRB~051117A}
\authorrunning{Goad et al.}
\maketitle

\section{Introduction}

The {\it Swift\/} Gamma-Ray Burst Explorer (Gehrels et~al. 2004), now
approaching the end of its second year of operations, is routinely observing
the prompt gamma-ray and early afterglow emission of Gamma-Ray Bursts (GRBs)
in the astrophysically important minutes to hours timescale after the onset of
the burst.  The greater sensitivity over previous gamma-ray missions of the
Burst Alert Telescope (hereafter BAT, Barthelmy 2004; Barthelmy et~al. 2005)
together with {\it Swift's\/} unique autonomous pointing and rapid slew
capability enables observations of the burst position in the narrow-field
instruments, the X-Ray Telescope (XRT, Burrows 2004; Burrows et~al. 2005) and
UltraViolet-Optical Telescope (UVOT, Roming et~al. 2005) on timescales of less
than 100~s, opening up to scrutiny a largely unexplored region of parameter
space.

Amongst {\it Swift's\/} many outstanding successes during the first two years
of operations, are the following notable highlights: the first localisation of
the afterglow emission from a short GRB (e.g. Gehrels et~al. 2005), the
detection of a very high redshift burst (e.g. Cusumano et~al. 2006), an
unexpected rapid decline phase in the prompt X-ray light-curves (Tagliaferri
et~al. 2005, Goad et~al. 2006), observations of frequent flaring activity in
the early X-ray light-curves of approximately half of all bursts (Burrows
et~al. 2005b; King et~al. 2005; Falcone et~al. 2006), and the intriguing
discovery of a burst which would have been classified as short by the Burst
and Transient Source Experiment (BATSE) on the Compton Gamma-Ray Observatory
(CGRO), but had a long soft tail in the BAT which lasted over 100~s
(e.g. Barthelmy et~al. 2005), to name but a few.  Indeed the many exciting new
discoveries by {\it Swift\/} have posed many challenges to existing
theoretical models of the prompt and early afterglow emission of GRBs.

In the standard fireball model (see eg. Zhang and M{\'e}sz{\'a}ros 2004 for a
thorough review), the core-collapse of a massive star produces a
relativistically expanding blast wave which decelerates as it impacts on the
surrounding interstellar medium.  The GRB spectrum is dominated by non-thermal
emission (either synchrotron or inverse Compton) from shock accelerated
relativistic electrons, which subsequently cool as the fireball expands,
causing the spectrum to shift toward lower energies.

 Large amplitude, short timescale fluctuations in both the prompt $\gamma$-ray
and early X-ray light-curves as observed by {\it Swift\/}, support the
suggestion that the GRB prompt emission arises from internal processes rather
than from external shocks (ie. before the blastwave has been decelerated by
the ambient medium). Possible scenarios include production via internal shocks
(Rees and M{\'e}sz{\'a}ros 1994, Kobayashi et~al. 1997), dissipation in strong
magnetic fields (Drenkhahn and Spruit 2002, Kumar et~al. 2007) or
Comptonization of the photospheric emission (Rees and M{\'e}sz{\'a}ros 2005).

Indeed, the early rapid declines in the X-ray light-curves, and the smooth
connection between the early X-ray light-curves and the tail of prompt
$\gamma$-ray emission (Barthelmy et al. 2005), suggests that the early X-ray
emission is simply an extension of the prompt $\gamma$-ray emission seen in
the initial explosion (Nousek et~al. 2006). The steep decay slopes ($f(t)
\propto t^{-3}$ or steeper) associated with the prompt X-ray emission
(e.g. Tagliaferri et~al 2005, Goad et~al. 2006) are also consistent with
internal processes.  If the central engine activity ceases abruptly, an
external observer will continue to see emission from increasingly large angles
($\theta > \Gamma^{-1}$) with respect to their line-of-sight, the so-called
"curvature effect" (e.g. Kumar and Panaitescu 2000; Dermer 2004; Fan and Wei
2005).  This steep phase is typically followed by a shallower decay phase
which is spectrally harder (e.g. Goad et~al. 2006, Nousek et~al. 2006,
Willingale et~al. 2007) and is thought to be associated with late time energy
injection (refreshed shocks) which suggests the central engine activity may
last for far longer than had previously been thought.  At later times the
light-curve may steepen to follow the canonical afterglow decay value
($\alpha\approx1.2-1.4$) observed in the decay of optical afterglow
light-curves.  Interestingly, jet-breaks accompanying the late time decay of
the X-ray afterglow are for the most part not detected by {\it Swift\/}
(eg. Willingale et~al. 2007).

At least one {\it Swift\/} GRB has an associated Supernovae
(GRB060218/SN2006aj, Campana et~al. 2006).  Given the observed apparent
association between some GRBs and SN (see eg.  Woosley and Bloom 2006, and
references therein), there has long been the expectation that emission-lines
would be detected in the early X-ray and optical afterglow spectra of GRBs.
Indeed, there have been numerous claims in the literature for the detection of
X-ray emission-lines in the X-ray spectra of GRB afterglows across a number of
X-ray platforms including Beppo-Sax, Chandra and XMM (eg. Piro et~al. 1999;
Piro et~al. 2000; Yoshida et al. 2001; Reeves et~al. 2002; Butler et~al. 2003;
Watson et~al. 2003), though in the majority of cases the detection
significance is modest (a few $\sigma$) at best. Line species identified
include FeK$\alpha$, with an energy consistent with the redshift of the host
galaxies, and/or lines associated with blue-shifted lighter elements of S, Si,
Ar, Mg and Ca.  Two possible scenarios have been proposed for the origin of
the X-ray emission-lines. In the first, post-burst energy injection is
reprocessed into line emission by material lying very close to the GRB
progenitor ($R\sim10^{13}$~cm) (Rees and M{\'e}sz{\'a}ros 2000). In this model
the lifetime for the line-emission is governed by the duration of the
post-burst injection phase.  Alternatively, the line emission may be formed by
reprocessing of both the prompt and early afterglow emission at relatively
large distances ($R\sim10^{16}$~cm, eg. Lazzati et~al. 2002; Reeves
et~al. 2002; Kumar and Narayan 2003). In this scenario, the lifetime of the
emission is determined by the size of the reprocessing region (see e.g. Gao
and Wei 2005 for an in-depth discussion of this model). Futhermore, the line
emission is expected to be detected within the first few hours of the initial
explosion.  To date despite early ($<$ few minutes) X-ray and optical
observations of more than 150 GRBs by {\it Swift\/}, there is as yet no
conclusive evidence for line emission in any of the afterglow spectra (see
e.g. Butler et~al. 2007; Hurkett et~al. 2007).

Here we report on {\it Swift\/} observations of the bright flaring burst
GRB~051117A, one of the brightest bursts observed with XRT.  The paper is
organised as follows. In \S2 we describe the data obtained by {\it
Swift\/}. In \S3 we present a detailed temporal and spectral analysis of the
BAT and early XRT data for this burst.  In \S4 we present observations of
GRB051117A in other bands.  In \S5 we place the observations in the context of
theoretical models of the GRB and afterglow emission. Our conclusions are
presented in \S6.

\section{Observations}

\subsection{GRB~051117A}

The {\it Swift\/} BAT triggered and located on board GRB~051117A
(trigger=164268) at 10:51:20 UT Nov 17th 2005 (Band et~al. 2005). The
spacecraft autonomously slewed to the burst location and began observations
with XRT at 10:53:07 UT. The XRT found a very bright, fading, uncatalogued
source at RA 15h 13m 33.8s, Dec +30d 52m 13.3s (J2000), with a positional
uncertainty of 3.4~arcsec (90\% containment, Goad et al. 2005b).  This
position which includes the latest XRT boresight correction (Moretti
et~al. 2006), lies 41 arcsec from the BAT on-board position and 3.8 arcsec
from the refined UVOT position (RA 15h 13m 34.1s, Dec +30d 52m 12.7s (J2000),
Holland 2005).


\subsubsection{BAT spectrum and light-curve of GRB~051117A}

The BAT light-curve of GRB~051117A is characterised by a long FRED-like (Fast
Rise Exponential Decay) peak at $T_{\rm 0}-15$~s lasting out to at least
$T_{\rm 0}+190$~s, with a hint of smaller emission peaks at $T_{\rm 0}+225$~s
and $T_{\rm 0}+350$~s (each of approximately 30~s in duration).  A simple
powerlaw fit to the BAT spectrum, covering the time interval $T_{\rm 0}-29$~s
to $T_{\rm 0}+157$~s, has a best-fit photon index $\Gamma= 1.83\pm0.07$, with
a fluence of $4.6\pm0.16\times10^{-6}$~erg~cm$^{-2}$ (all values in the
15-150~keV band). The peak flux in a 1~s wide window starting at $T_{\rm
0}+2.47$~s is $0.93\pm0.17$~ph~cm$^{-2}$~s$^{-1}$ (quoted errors are 90\%
confidence on 1 interesting parameter).  Relative to other {\it Swift\/}
detected GRBs, GRB051117A is rather unremarkable, with a BAT fluence placing
it at the mean of the brightness distribution, while its spectral
characteristics place it at the soft end of the burst population. In fact
based on Lamb's  definition (Lamb et~al. 2005),  GRB051117A qualifies as an
X-ray Rich Burst.

\subsubsection{XRT observations of GRB~051117A}

The XRT began observing GRB~051117A 107~s after the BAT trigger. The on-board
software located a very bright uncatalogued X-ray source in a single 0.1~s
Image Mode (IM) frame, with an initial flux estimate of
$1.0\times10^{-8}$~erg~cm$^{-2}$~s$^{-1}$ in the 0.2-10~keV band (this assumes
an unabsorbed powerlaw model with photon index $\Gamma=2.0$).

GRB~051117A is amongst the brightest (in terms of observed counts) early X-ray
light-curves yet observed by {\it Swift\/}. Indeed, the source was so bright
in XRT ($>$200 ct~s$^{-1}$ in the IM frame) and the initial rate of decline so
slow ($\alpha=0.77$, where $f(t)\propto t^{-\alpha}$), that the source
remained in Windowed Timing (WT) mode for the whole of the first orbit,
providing an unprecedented 1770~s of WT mode data in the first 2 orbits. XRT
switched into Photon Counting (PC) mode 170~s into the 2nd orbit of
observations (5187~s after the BAT trigger). We note that even in orbit 2, the
PC mode data are piled-up, which is highly unusual for data taken at such a
late time.  For later orbits, XRT remained in PC mode and followed the source
until it faded into the background, with a total on source exposure time of
$\approx 430$~ks (see Table~1 for details of the observations).

The WT and PC mode event lists were processed using the standard XRT data
reduction software {\bf xrtpipeline} version 0.9.4, within FTOOLS v6.0.3,
screening for hot and flickering pixels, bad columns, and selecting event
grades 0-12 for PC mode data, 0-2 for WT mode data.  Table~1 lists the
name, start and stop times (in mission elapsed time, MET) and total on source
exposure time of each observing segment.  Observations of GRB~051117A were
continued until Dec 11th 2005, yielding 24 days of coverage.  Due to the close
proximity of GRB~051117A to the Sun direction, there were no extensive
ground-based follow up observations of this source in the immediate days to
weeks after the burst.  Fortunately, the location of GRB~051117A was in a
favourable sky position for XRT observations\footnote{Failure of the XRT
Thermo Electric Cooler (TEC) during the calibration and verification phase of
the mission has introduced a change to the mission operations, with an
additional workload on the observation planners, with the requirement that GRB
observations must maintain the XRT temperature below -47 degrees Celsius. This
essentially means choosing space-craft roll angles which keep the radiator
pointed away from the Earth and other non-favourable directions. For higher
temperatures hot and flickering pixels increase substantially and can result
in frequent and undesirable mode switching.}, and the CCD temperature remained
cool for the duration of our observations.

\begin{figure}
\resizebox{\hsize}{!}{\includegraphics[angle=0,width=8cm]{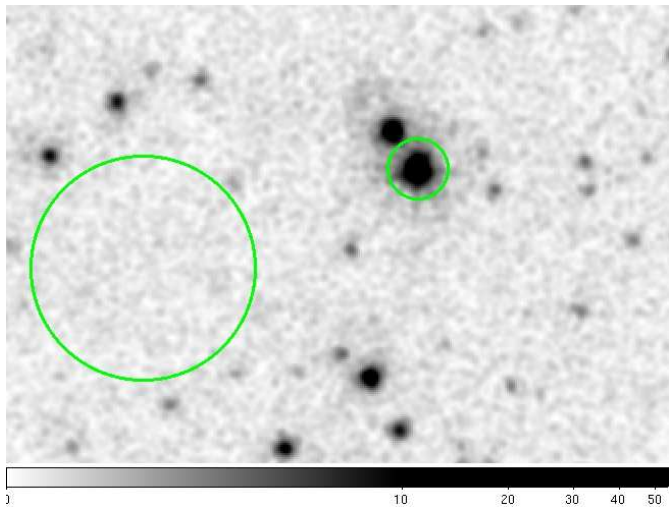}}
\caption{A deep, smoothed, XRT image formed from summing the cleaned event
lists for all PC mode observations, excluding those taken during orbit 2 which
are affected by pile-up. The chosen 15 pixel source and 55 pixel background
extraction regions are indicated by the small and large circles
respectively. The total exposure time for the centre of this field is 430
ks. 99 sources were detected at greater than $5\sigma$.}
\label{GRB051117a_batlc}
\end{figure}

In Figure~1 we show the central few arcminutes of a deep ($\sim$430 ks) XRT
observation of the field of GRB~051117A, formed from the summed cleaned PC
event lists from orbit 3 onwards.  There are 99 sources in the field detected
at $> 5$ sigma.  Of particular note is the bright stellar source at RA 15h 13m
36.4, Dec +30d 52m 57.9~s (J2000)
which lies only 54 arcsec from the XRT position of the GRB. Due to the close
proximity of this source to the GRB we used a smaller than typical source
extraction region for PC mode data, a circular aperture of outer radius 15
pixels ($\equiv35.4$ arcsecs), compared to the default 20 pixel radius region
used in the standard pipeline analysis. Similarly, for the WT mode data we use
a 30 pixel wide box centred on the source (c.f. the default 40 pixel wide
box). Due to the large number of background sources, the background region had
to be chosen with care. For the PC mode data we choose a circular background
region of 55 pixel radius located at RA 15h 13m 58.3s Dec +30d 50m 22.3s
(J2000, uncorrected for boresight), and which appears to be free from
contamination by background sources (see Fig~1). For the WT mode data we
choose a background region with the same size as the source region and at a
radial distance that places it well outside of the location of the nearby
bright star. Finally, increased X-ray flaring activity observed in the X-ray
light-curve of the nearby star $1.8\times10^{6}$~s after the burst, forced us
to use an even smaller 10 pixel wide extraction region, in order to minimise
contamination of the XRT light-curve at the lowest count rates.  Our chosen
source and background regions (small and large circles respectively,
together with a number of other sources detected in this field are shown in
Fig~1.

We note that the WT mode data taken during the first and second orbits falls
upon the bad columns, which appeared following a micro-meteorite impact on 27th
May 2005 (Abbey et~al. 2005). Comparison of the XRT PSF with and without
the bad columns shows that the WT data is affected at the 1.4\% level. The WT
mode data has been scaled accordingly.  Finally, as mentioned earlier, the PC
mode data taken during orbit 2 are affected by pile-up. To account for this we
use an annular source extraction region of inner radius 4 pixels, and outer
radius 15 pixels, for the 2nd orbit of data only.

\begin{figure}
\resizebox{\hsize}{!}{\includegraphics[angle=270,width=8cm]{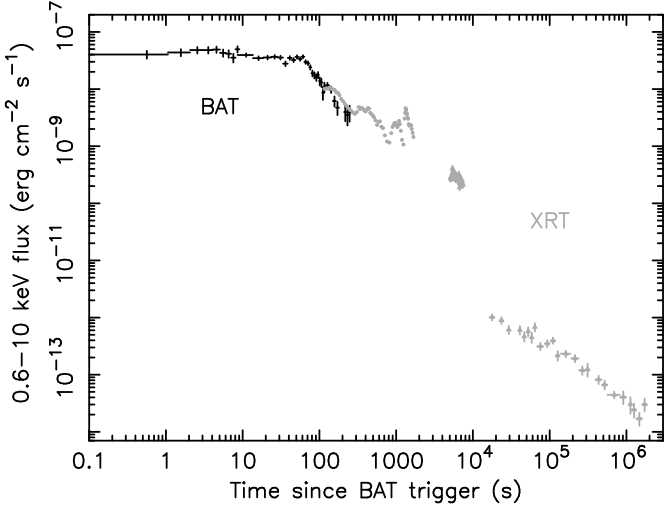}}
\caption{The combined BAT 15-350~keV mask-weighted light-curve and XRT WT mode
(orbits 1 and 2) and PC mode (orbits 2 and later) light-curve. The BAT
count-rate light-curve has been converted into the 0.6-10.0 keV bandpass
using the combined spectral fit to the BAT-XRT data taken in the overlap
region. Similarly XRT count rate data from orbits 2 and later have been
converted into flux units using the mean spectral fits to the XRT data taken
during these 2 epochs.}
\label{bat_xrt_lc}
\end{figure}

\subsection{Combined BAT and XRT light-curve}

The combined BAT/XRT light-curve of GRB~051117A is shown in
Fig~\ref{bat_xrt_lc}. The BAT light-curve has been constructed by
extrapolating the BAT data into the 0.6-10~keV XRT band, using the mean photon
index of the combined fit to the BAT and WT mode XRT data in the 190~s overlap
region.  The XRT data were converted into flux units using the mean spectral
fits to the 0.6-10~keV band during the first orbit (WT mode data), 2nd orbit
(PC mode data), and late-time PC mode data (orbit 3 onwards).  For clarity the
1st orbit WT mode data have been binned to $> 1000$ ct/bin.  Orbit 2 (WT and
PC mode data) data have been binned to $>100$~ct/bin. For orbit 3 and later,
data have been binned to a minimum 20 ct/bin at early times, and for
$>3\sigma$ detections at late times. The estimated mean background count rate
over $\approx400$~ks of observations within a 15 pixel radius circular
aperture is $3.76\pm0.08\times10^{-4}$~ct~s$^{-1}$.  An extrapolation of the
late-time BAT light-curve appears to join smoothly with the XRT data taken at
the beginning of orbit 3, suggesting that the flaring activity may have lasted
at least until the end of the orbit 2 observations (7.5 ks after the BAT
trigger).

The XRT count rate light-curve can be approximated by a series of broken
powerlaws with superposed flares at early times.  The initial decline appears
shallow, with an underlying slope of $\alpha=+0.77^{+0.08}_{-0.06}$, and
displays numerous, relatively low amplitude short-lived flares. This
relatively shallow early decay, suggests that the early X-ray light-curve
behaviour is dominated by flaring activity.  Indeed a simple extrapolation of
the late time BAT light-curve to the late time XRT light-curve requires an
underlying decay slope of $t^{-3}$, more typical of the fast X-ray decay
slopes observed in many GRBs at early times (e.g. Tagliaferri et al. 2005).

The light-curve breaks sharply for $T_{0}+7450 < t < T_{0}+16500$~s with
$\alpha>5$, though the exact time of the break is undefined, before flattening
again, with a slope of $\alpha=0.66\pm 0.09$. At late times, the light-curve
breaks once more to a steeper slope with $\alpha=1.1^{+0.16}_{-0.14}$, at
$t_{\break} = T_{0}+168440^{+97360}_{-80350}$~s.  The break in the light-curve
at late times is highly significant. A fit to the late time data using a
broken powerlaw as opposed to a single powerlaw yields $\Delta \chi^{2} = 18$
for 2 fewer degrees of freedom.

\begin{figure}
\resizebox{\hsize}{!}{\includegraphics[angle=270,width=8cm]{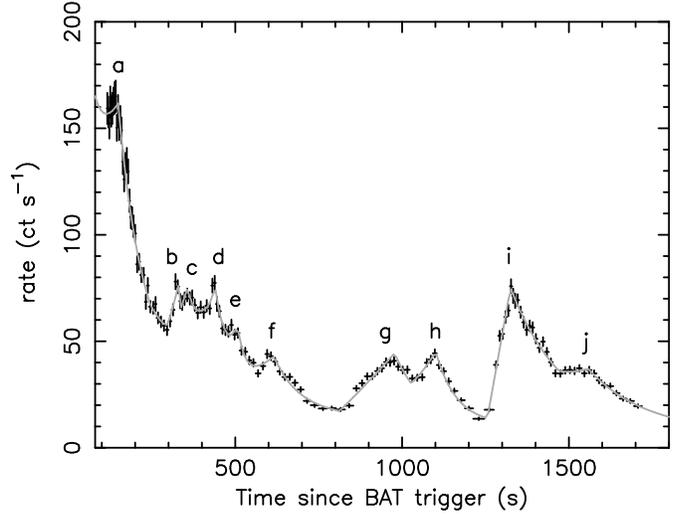}}
\caption{Model fit to early X-ray light-curve. The light-curve is well-fit by
a single powerlaw with slope $-0.77$, with superposed FREDs  (labelled a-j, see
also Table 2) spanning a range in strength and duration, $\chi^{2} = 390$ for
376.}
\label{fig_flares}
\end{figure}

\subsection{Timing characteristics of the prompt X-ray light-curve}

The early X-ray light-curve of GRB~051117A is characterised by a moderately
slow decline with random, superposed flares of varying strength and duration.
Visually the shape of the flares, and in particular the last flare in the
first orbit of data, appear consistent with a fast-rise followed by an
exponential decay (FRED-like), a form commonly observed in the prompt
gamma-ray emission of many GRBs including GRB~051117A.  In Table~2 we give the
results of fitting the X-ray light-curve with a single powerlaw decay slope,
with slope $0.77$ with a series of superposed FRED-like components, indicating
the rise times and e-folding times for each of the individual flares. Note
that the powerlaw is required to match the WT and PC mode data taken during
orbit 2 (see Fig~\ref{bat_xrt_lc}).  A power spectral analysis of the first
orbit XRT data confirms the shot-noise like nature of the emission process
(see eg. Lehto 1989), with most of the power emitted on the longest
timescales, and declining smoothly with a powerlaw slope of $\approx$2.6,
before flattening to the noise level on timescales shorter than 20~s.

\section{The X-ray spectral properties of GRB~051117A}



 Before {\it Swift\/}, prior evidence for excess absorption from GRB X-ray
spectra is sketchy at best (see Stratta et~al. 2004, De Luca et~al. 2005,
Gendre et~al. 2005).  More recently, in an analysis of 17 Swift detected GRBs
(Campana et~al. 2006) found evidence for an observed column density above the
Galactic value in $\sim60$\% of cases. Campana et~al. attribute the excess
columns to an origin for long GRBs within dense molecular clouds (see also
Galama and Wijers 2001).  We note that if the absorbing material is close to
the GRB, the high radiation field can photo-ionise the surrounding material
out to several parsecs, reducing the amount of low-energy aborption with time
(Lazzati and Perna 2002).  The detection of excess Nh, variable or otherwise,
is important as it provides strong supporting evidence for massive star
progenitors and the association of long GRBs with star-forming regions
(Fruchter et~al. 1999, Prochaska et~al. 2004). Below we present a detailed
spectral X-ray analysis of the early and late time X-ray specta for
GRB051117A.

\subsection{Joint BAT/XRT spectrum}

\begin{figure}
\resizebox{\hsize}{!}{\includegraphics[angle=270,width=8cm]{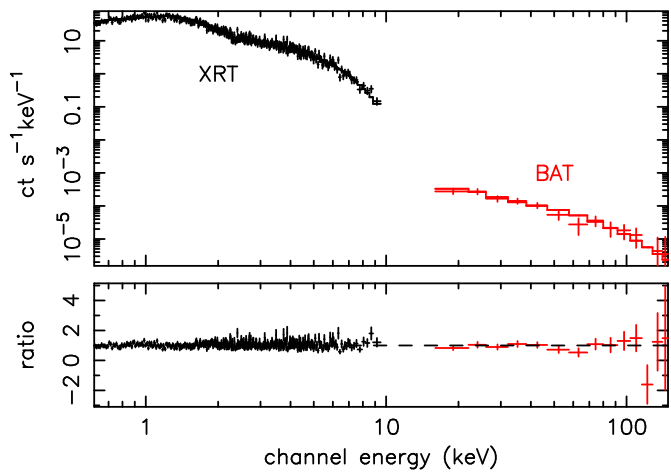}}
\caption{A fit to the combined BAT-XRT spectrum covering the time interval
$T_{\rm 0}+113$~s to $T_{\rm 0}+303$~s. The data are well fit by a Band
function, with Nh$_{\rm excess}=1.6\pm0.6\times 10^{21}$~cm$^{-2}$,
$\alpha=-0.43^{+0.3}_{-0.84}$, $\beta=-1.95\pm0.04$, and $E_{\rm
  break}=1.0^{+0.7}_{-\infty}$~keV, $\chi^{2}=414$ for 404 dof.}
\label{bat_xrt}
\end{figure}

The long duration of this GRB provides a 190~s observation window (covering
the time interval $T_{\rm 0}+113$~s to $T_{\rm 0}+$303~s) for which we have
significant counts in both the BAT and XRT.  In Figure~\ref{bat_xrt} we show
the joint BAT/XRT spectrum extracted over this time interval. The joint
BAT/XRT spectrum can be well fit with a single absorbed powerlaw with Nh$_{\rm
excess}=2.3\times 10^{21}$~cm$^{-2}$, photon index $\Gamma=2.0$,
$\chi^{2}=421$ for 409 dof.  A Band function (Band et~al. 1993) with excess
absorption, also provides an acceptable fit to the data, with Nh$_{\rm
excess}=1.6\pm0.6\times 10^{21}$~cm$^{-2}$, $\alpha=-0.43^{+0.3}_{-0.84}$,
$\beta=-1.95\pm0.04$, and $E_{\rm break}=1.0^{+0.7}_{-\infty}$~keV, $\chi^{2}=414$ for
404 dof. We note that while excess Nh is required by the Band function
  model fit ($\Delta \chi^{2}=52$ for one less dof), we obtain only an upper limit to
the break energy with this model.

\subsection{The prompt X-ray spectrum}

\begin{figure}
\resizebox{\hsize}{!}{\includegraphics[angle=270,width=8cm]{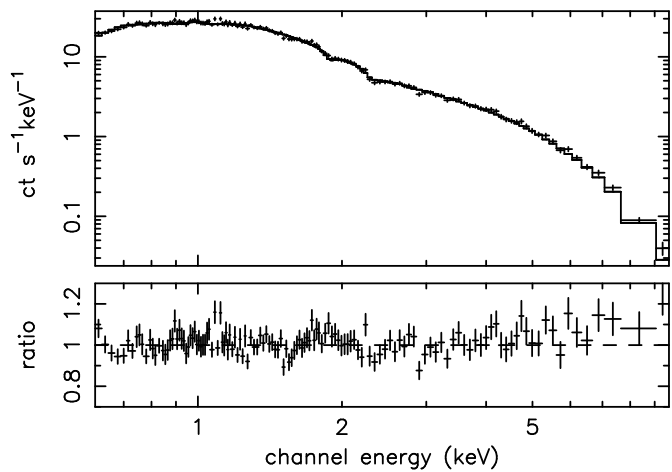}}
\caption{WT mode spectrum formed from the first orbit of WT mode data
(1605~s).  The spectrum is best fit by an absorbed broken powerlaw model, with
Nh above the Galactic value, with $\Gamma_{1}=1.77\pm0.23$,
$\Gamma_{2}=2.19\pm0.03$, and break energy $E_{\rm break} =
1.04^{+0.08}_{-0.13}$ (all errors 90\% confidence), $\chi^{2}=511$ for 531
dof (see text for details).}
\label{wt_1storb}
\end{figure}

The high observed count rate and slow rate of decline during the first orbit
of XRT WT mode data, allows us to construct amongst the highest quality S/N
spectrum of any burst yet observed by {\it Swift\/} comparable to the bright
bursts GRB061121 (Page et~al. 2006) and GRB060124 (Holland et~al. 2006). In
Figure~\ref{wt_1storb} we show a model fit to the mean WT mode spectrum
accumulated during the first orbit (1605~s of data). Fitting a single absorbed
powerlaw to the 0.6-10.0~keV band, with Nh tied at the Galactic value of
$1.82\times 10^{20}$~cm$^{-2}$ is a very poor description of the data,
$\chi^{2} = 2777$ for 534 dof. Allowing for excess Nh provides a significant
improvement, with a best-fit model of $\Gamma=2.23\pm0.02$, ${\rm Nh}_{\rm
excess}=2.11\pm0.01\times10^{21}$~cm$^{-2}$, with $\chi^{2}=523$ for 533
degrees of freedom, all errors 90\% confidence on 1 interesting parameter.  An
absorbed broken powerlaw model, with Nh above the Galactic value, provides an
improved fit to the data, with a best-fit $\Gamma_{1}=1.77\pm0.23$,
$\Gamma_{2}=2.19\pm0.03$, and break energy $E_{\rm break} =
1.04^{+0.08}_{-0.13}$, $\chi^{2}=511$ for 531 dof.

We have searched for the possible existence of emission-lines in the WT mode
spectrum taken during the first orbit of observations.  At near the peak of
the effective area (1~keV), we can rule out the presence of narrow spectral
features ($\sigma=100$~eV) with EW less than 15~eV at greater than 3$\sigma$
confidence. Broader spectral features cannot be excluded by the data, due to
uncertainties in the spectral calibration.

\subsection{Global spectral properties}

To investigate the long term spectral evolution of the prompt and early
afterglow emission of GRB~051117A, we have formed average spectra  (see Figure
6) covering the 1st orbit (WT data only, 1605~s), 2nd orbit [(both WT (163~s)
and PC mode data (2313~s)] and the late time PC mode data (orbit 3 onwards,
411~ks).  A single absorbed power-law fit to each of the four spectra with
excess Nh above the Galactic value of $1.82\times10^{20}$~cm$^{-2}$ is a good
description of the data, with $\Gamma=2.24\pm0.02$, ${\rm Nh}_{\rm
excess}=2.1\times 10^{21}$~cm$^{-2}$, $\chi^{2}=708$ for 679 dof. Untying
$\Gamma$ between each of the 4 segments, provides a significant improvement,
$\Delta \chi^{2}=42$, F-statistic 14.3, null hypothesis $4.7\times10^{-9}$. We
note that keeping $\Gamma$ tied and untying Nh provides an equally good
description of the data, though excess Nh is consistent with being constant
within the errors.
 
Finally, we tried to fit an absorbed broken powerlaw to each of the segments,
with $\Gamma$ tied pre- and post-break and excess Nh of $1.3\times
10^{21}$~cm$^{-2}$ tied between the segments. This model is only marginally
worse, with $\chi^{2}=669$ for 674 dof.  Untying $\Gamma$ between the spectra,
significantly improves the fit $\chi^{2}=645$ for 668 dof (see
Table~3 for details). With such good quality data, the lack of evidence for a
variable Nh suggests that the absorbing column may not be associated 
with the immediate GRB environment.

\begin{figure}
\resizebox{\hsize}{!}{\includegraphics[angle=270,width=8cm]{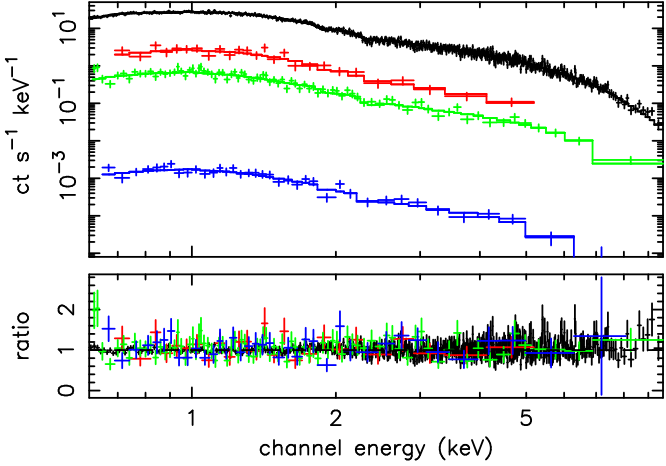}}
\caption{WT mode (black and red, 1st and 2nd orbit respectively), PC
mode(green 2nd orbit only, blue orbit 3 and later) spectra (0.6-10 keV) for
GRB~051117A. The data are well fit by an absorbed powerlaw, with an excess Nh
above the Galactic value of $1.82\times10^{20}$~cm$^{-2}$ of
$2.1\times10^{21}$~cm$^{-2}$, and $\Gamma$ increasing through each of the 4
segments, $\chi^{2}=666$ for 676 dof.}
\label{glob_spec}
\end{figure}

\subsection{Spectral characteristics of the early X-ray light-curve}

\begin{figure}
\resizebox{\hsize}{!}{\includegraphics[angle=270,width=8cm]{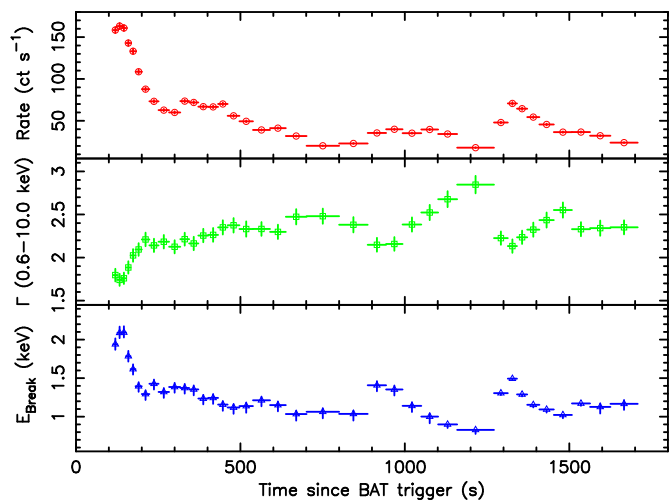}}
\caption{Upper panel - The 0.6-10.0~keV WT mode light-curve of the 1st orbit
data, binned to 2000~ct~bin$^{-1}$.  Middle panel - The temporal variation in
photon index $\Gamma$, derived from a simultaneous fit to 37 spectra, assuming
a simple absorbed powerlaw, with excess Nh tied between each of the segments
and $\Gamma$ allowed to freely vary. Lower panel - The variation in break
energy, $E_{\rm break}$, for the same spectra, adopting a broken powerlaw
spectral model with $\Gamma$ pre- and post-break tied between the spectra. In
this model excess Nh is not included  (see text for details).} 
\label{segs}
\end{figure}

The superb quality, ie. high observed count rate and relatively slow temporal
decline, exhibited by the early X-ray light-curve of GRB~051117A during the
first orbit affords a detailed temporal analysis of the X-ray spectral
evolution for this source. For this analysis, the early X-ray light-curve was
divided into segments containing approximately 2000~ct~bin$^{-1}$, and spectra
extracted for each segment.  This binning scheme represents the best
compromise between spectra with approximately equivalent signal-to-noise, and
high temporal resolution, and allows us to investigate the X-ray spectral
evolution of this source in a uniform manner.  The binned light-curve composed
of 37 segments of data is shown in Figure~\ref{segs} (top panel). The early
XRT light curve is characterised by a relatively slow temporal decline
($\alpha=0.77$) with superposed flares of varying strength and duration.  For
each of the 37 segments of WT mode data we extract spectra from the cleaned
event list, using the standard grade selection (grade 0-2 for WT mode), and
using a 30 pixel wide window centred on the source position.

\begin{figure}
\resizebox{\hsize}{!}{\includegraphics[angle=270,width=8cm]{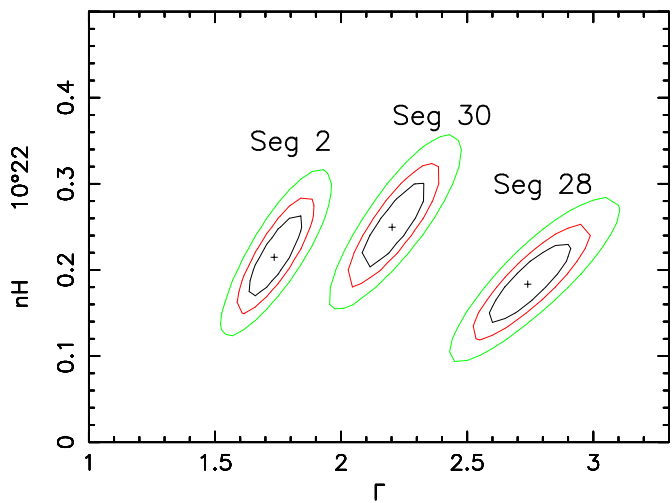}}
\caption{Variation of excess Nh and photon index $\Gamma$ for segments 2, 28
  and 30.}
\label{excess_2}
\end{figure}

A simultaneous fit to the 37 spectra over the 0.6-10.0 keV band, using a
simple absorbed powerlaw, with Nh fixed at the Galactic value of
$1.82\times10^{20}$~cm$^{-2}$, excess absorption of
$2.1\times10^{21}$~cm$^{-2}$ with $\Gamma$ and the normalisation untied
between segments, provides an acceptable fit $\chi^{2} = 2470$ for 2648
dof. Figure~\ref{segs} (middle panel) shows the derived photon indices for
each time segment (all errors 90\% confidence). Similarly, an absorbed broken
powerlaw fit to the spectra, with Nh fixed at the Galactic value, the photon
index pre-break tied together, the photon index post-break tied together, and
the break energy $E_{\rm break}$ and normalisation untied, also provides an
acceptable fit to the data $\chi^{2}=2593$ for 2647 dof, without requiring
excess Nh. Including excess Nh only moderately improves the fit
$\Delta\chi^2=22$. The variation in break energy for each of the 37 segments
for the broken powerlaw model without excess Nh as a function of time is shown
in Fig~\ref{segs} (lower panel). We checked for the presence of variable
excess NH in our data, by allowing Nh to freely vary between
segments. However, although there is clear evidence for variable $\Gamma$, we
find no evidence for variable Nh. Figure~\ref{excess_2} shows the 68\%,
95\% and 99\% confidence contours for $\Gamma$ versus Nh taken from 3 data
segments spanning the full range in source count rate.
\footnote{We have checked that temperature dependent variable gain and bias
offsets, which can produce $\sim20$~eV shifts in the energy scale do not
effect our conclusions.}

\begin{figure}
\resizebox{\hsize}{!}{\includegraphics[angle=0,width=8cm]{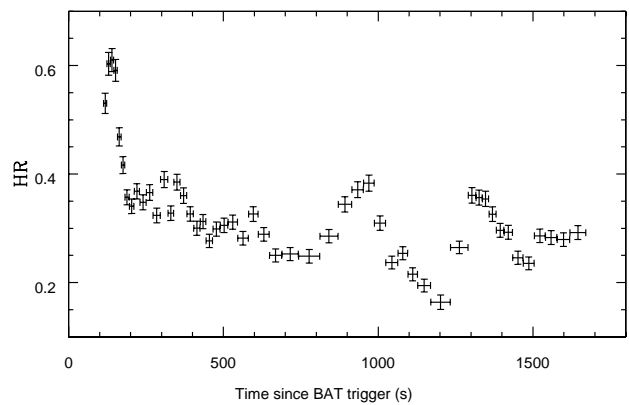}}
\caption{Hardness ratio HR=H/S, where HR is the ratio
of counts in the $2.0-10.0/0.2-2.0$ ~keV band.}
\label{hr}
\end{figure}

\begin{figure}
\resizebox{\hsize}{!}{\includegraphics[angle=270,width=8cm]{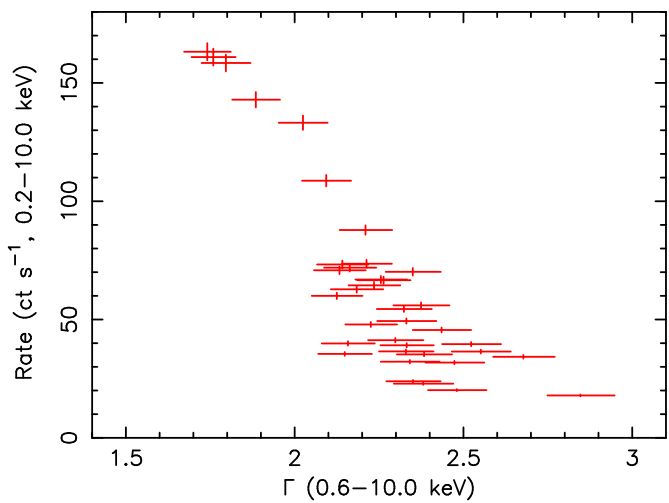}}
\caption{Photon index as a function of count-rate. This confirms that the
  spectra are harder when brighter}
\label{inten}
\end{figure}

Visual inspection of Fig~\ref{segs}, panels 1 and 2, shows that there exists a
strong anti-correlation between the source intensity and photon index, such
that the spectrum is much harder when the source is brighter. This can be seen
more clearly in Figure~\ref{hr} where we plot the hardness, HR, here defined
as the ratio of the counts in the 2.0-10.0~keV band relative to the
0.2-2.0~keV band, as a function of time. Source intensity and hardness ratio
are found to be highly correlated with a Pearson's linear correlation
coefficient $r=0.85$ with a probability $P$ of no correlation which is
vanishingly small.  In Fig~\ref{inten} we show the source intensity (0.6-10
keV) as a function of photon index $\Gamma$. This again shows the source is
spectrally harder when brighter, though we note that the exact correspondence 
between the two parameters is not simply one to one. Moreover, the
variation in $\Gamma$ with source intensity appears to shows signs of
hysteresis. That is, the variation in $\Gamma$ with source intensity for each
flare, appears to follow similar tracks. The observed spectral behaviour is
entirely consistent with the gamma-ray spectral evolution seen in BATSE GRBs
(e.g. Ford et~al 1995; Bhat et~al 1994).

\begin{figure}
\resizebox{\hsize}{!}{\includegraphics[angle=0,width=8cm]{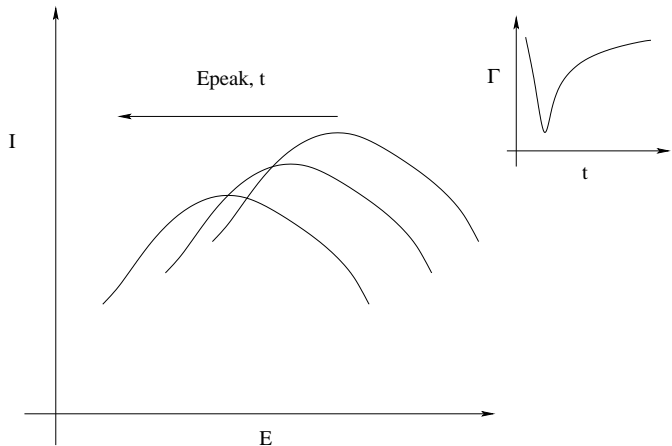}}
\caption{Cartoon model indicating the variation in photon index $\Gamma$
(inset), for a spectrum consisting of a broken powerlaw, as the break-energy
moves to lower energies.}
\label{cartoon}
\end{figure}

\begin{figure}
\resizebox{\hsize}{!}{\includegraphics[angle=270,width=8cm]{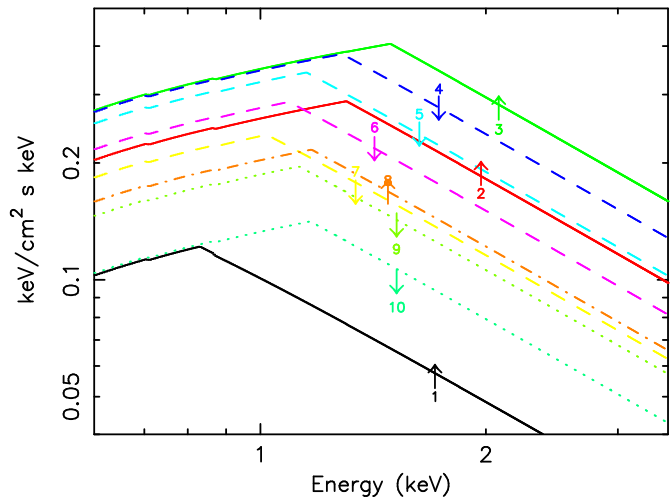}}
\caption{Model fits to the X-ray spectra extracted from the last observed
flaring episode of first orbit of data (10 spectra of approx 2000 ct, numbered
sequentially from 1--10). Here we
model the data as an absorbed broken powerlaw, with $\Gamma$ pre-break tied
together, $\Gamma$ post-break tied together, and excess Nh tied together. The
general trend is for an increase in break energy as the source intensity
increases (fits 1-3), and a decrease in break energy as the source intensity
decreases. We note however, that while clearly correlated, 
there is no one-one correspondence between break-energy and source intensity.}
\label{cartoon_fits}
\end{figure}

Both Figure~7 (middle panel) and Figure~9, indicate that the hardening of the
spectrum coincides with each new flaring episode. The spectrum is then
observed to soften gradually (this can be seen either in terms of an increase
in the Photon index or a decrease in HR), as the X-ray intensity fades (see
also Figure~\ref{inten}). The observed strong correlation between source
intensity and spectral shape suggests a common origin for each of the flares.
The simplest mechanism which can account for the observed temporal behaviour,
is one in which the overall spectral shape remains constant, but the
characteristic break energy E$_{\rm break}$ varies with source intensity (see
also Crider et~al. 1997). Over the limited band-pass of the XRT, a small
increase in the break-energy will harden the spectrum, while a small decrease
in break-energy will soften the spectrum. In Figure~\ref{cartoon} we show a
cartoon illustrating this behaviour, and (inset) the resultant variation in
photon index $\Gamma$. In Figure\ref{cartoon_fits} we show the model fits to
the X-ray spectra for each of the segments covering the last flare (segments
28-37).  We note that the peak of the Band function fit (1 keV) to the joint
BAT-XRT data taken during the time interval $T_{\rm 0}+113$~s$ < t < T_{\rm
0}+303$~s does indeed lie within the XRT bandpass, adding strong support to
this simple picture.

\section{Observations of GRB~051117A in other bands}

{\it Swift\/} UVOT began observing GRB~051117A at 10:53:10 UT, 111 seconds
after the BAT trigger. A faint uncatalogued source was detected in a 200~s 
V-band image, with a magnitude of $V = 20.01\pm0.46$ (1-$\sigma$, statistical,
uncorrected). The extinction in this direction is $A_{V} = 0.08$ (Band et
al. 2005, Holland et~al. 2005).  In a 50~s UVOT white-light observation
starting 531 s after the BAT trigger the source had a magnitude of
$19.19\pm0.16$ (1-$\sigma$, statistical). 3.5 days after the burst the source
had faded to a white-light magnitude of approximately 21.5 (see Fig~\ref{uvlc}
and Table~4 for a detailed description of the UVOT
observations). Interestingly, there is weak evidence that the source is
extended, and which may be associated with the host galaxy of GRB~051117A.
During the fast decay observed in XRT $\approx10^{4}$ s after the burst, the
V-band magnitude remained remarkably constant. The optical flux declines
rapidly (as seen in the white light filter) $3\times 10^{4}$~s after the BAT
trigger.

In the interval 1000~s to 300~ks after the BAT trigger, the source intensity
(in both V and White light) decreased by 3 magnitudes, equivalent to a
powerlaw decay rate of $\alpha=0.5$.  Since this interval covers the sharp
break in the XRT light-curve, the optical light-curve has a much shallower
decay than the X-ray light-curve.
\begin{figure}
\resizebox{\hsize}{!}{\includegraphics[angle=270,width=8cm]{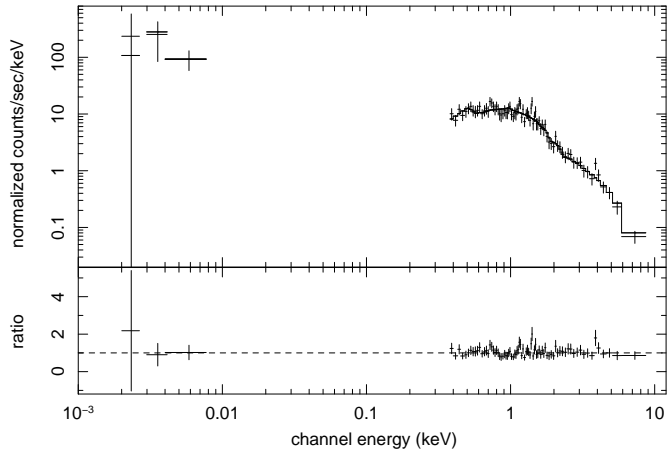}}
\caption{Broad-band spectral energy distribution (SED) of GRB051117A. This SED
was compiled at a logarithmic midpoint time of 750~s after the BAT trigger
(see text for details).}
\label{uvxray}
\end{figure}
We compiled a spectral energy distribution for the UVOT and XRT data at 750
seconds after the BAT trigger to investigate absorption and spectral shape in
the UV and optical data (Figure 13). Inclusion of the XRT data ensures a more
accurate fit to the underlying power law, but we note that near infrared data,
not available for this source, are also required for a precise determination
of the underlying continuum.  We find that a single power law cannot
adequately fit the SED, and a broken power law is required. We obtain a good
fit with a pre-break slope of $\Gamma\sim1$, breaking in between the optical
and X-ray data to $\Gamma=2.20\pm0.16$, consistent with the slope measured
from the X-ray data alone.  This spectral break cannot be identified as a
cooling break due to the large errors on the optical/UV spectral slope, but we
note that in the case of a cooling break $\beta_2 = \beta_1 + \frac{1}{2}$
(where $\beta = \Gamma - 1$) which is not found in the best-fitting model to
this SED.
 
Assuming a redshift of $z=0.73$, as estimated in Section~5.1, we clearly
detect excess absorption in both the optical and the X-ray regimes. We find
$E(B-V) \sim 0.12$ for either a Large or Small Magellanic Cloud extinction law
(Pei 1992) and $N_{\rm H} \sim 1.5 \times 10^{21}$~cm$^{-2}$, again consistent
with fits to the X-ray data alone.

\begin{figure}
\resizebox{\hsize}{!}{\includegraphics[angle=0,width=8cm]{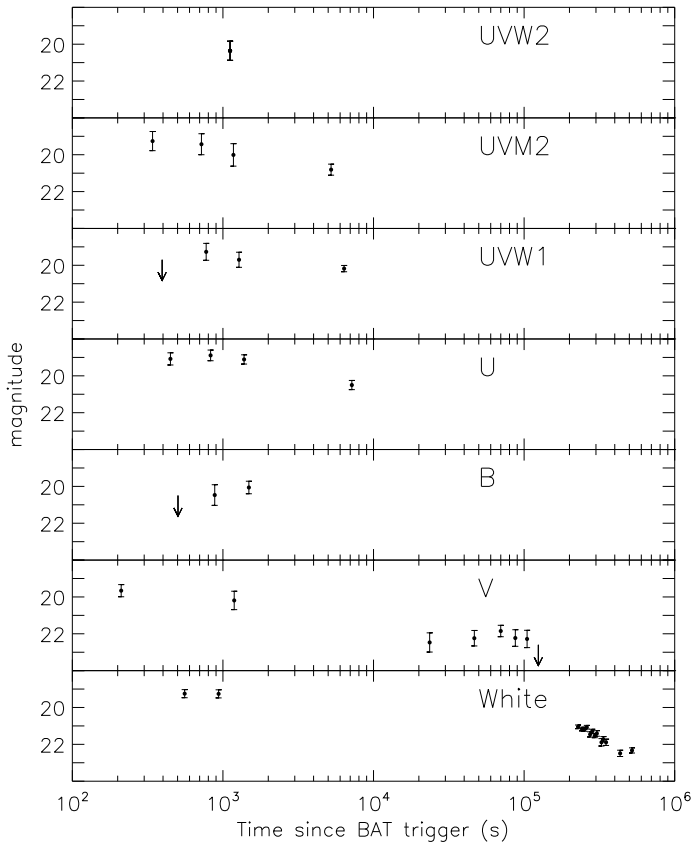}}
\caption{UVOT ultraviolet (UW2,UVM2,UW1), and optical (U,B,V, and white) light-curves of GRB051117A.}
\label{uvlc}
\end{figure}

Radio observations starting approximately 8 hours after the BAT trigger, with
the Very Large Array operating at 8.5~GHz did not detect an afterglow at
greater than 3-$\sigma$, for an rms noise level of 56~$\mu$Jy (Frail, D. 2005,
GCN 4292).

\section{Discussion}

The standard afterglow model of a spherical blast wave expanding uniformly
into a spherically symmetric ambient medium with a smooth density profile,
predicts smooth afterglow light-curves. By contrast the early X-ray decay
light-curve of GRB~051117A shows a highly complex structure which can be
modelled as a superposition of FRED-like shots on a smooth underlying powerlaw
decay.  Complex flaring behaviour in the early X-ray light-curve appears to be
a relatively common feature, being seen in approximately half of all GRBs
observed by {\it Swift\/} (Burrows et~al 2005, Falcone et~al. 2006). Suggested
origins for the appearance of flares include : (i) density fluctuations in the
medium into which the blast-wave expands, (ii) patchy shells, (iii) refreshed
shocks, and (iv) late time engine activity (Ioka, Kobayashi and Zhang 2005).
As noted by Ioka, Kobayashi and Zhang (2005), the relative fluctuation
amplitudes ($\Delta F_{\nu}/F_{\nu}$) and timescales ($\Delta t/t$) of
individual flares can be used to rule out some of these models.  For
GRB051117A, even the longest flares have timing properties which are
inconsistent with refreshed shocks ($\Delta t/t < 1/4$).  The patchy shell
model also fails, as $\Delta t < t$ in all cases.  While the low-amplitude
flares have timescales ($\Delta t/t \ll 0.01$) which are marginally consistent
with an origin due to small-scale density fluctuations in the ambient medium
(in the off-axis case), density fluctuations cannot explain either the
observed shape of the flares (flares due to density enhancements are typically
characterised by a slow rise to peak, e.g. Nakar and Granot 2006) or the
origin of the longer duration flares. Furthermore, invoking two distinct
mechanisms for flaring behaviour for flares which, in all other respects,
exhibit similar spectral and timing, characteristics, is somewhat
unsatisfactory.  Though it is possible that a variety of mechanisms may be
operating at the same time, the simplest explanation is that the observed
variations are in fact indicative of late time engine activity in this burst
(see also Lazzati and Perna 2007 for a full discussion on the origin of flares
in GRBs, and Guetta et~al. 2006 for an in depth analysis of X-ray flaring
activity in {\it Swift\/} GRB observations.). The close agreement between the
BAT and XRT light-curves in the overlap region, and the apparent connection of
the late-time BAT light-curve and XRT light-curve from orbit 3, strongly
supports this hypothesis. Moreover, the power spectrum of the early X-ray
light-curve, after detrending the data to remove the underlying steep decay,
is consistent with a shot noise process, which implies by definition a range of
amplitudes and timescales for individual flares (see Fig~\ref{fig_flares}, and
Table~2).

If the flaring activity is indeed related to late-time energy injection
(i.e. continued activity of the central engine), then for GRB~051117A the
engine possibly remains active for far longer than previously supposed.  The
steep decline for times $T_{0}+7450 < t < T_{0}+16500$ is most-likely due to
the curvature effect. If so, then this represents amongst the latest
detections of the curvature effect (even after allowing for correction to the
rest-frame) of any GRB, comparable to that observed in GRB050904 (Cummings
et~al. 2005), and GRB070110 (Krimm et~al. 2007).

\subsection{Redshift and luminosity}

For GRB~051117A there is currently no redshift estimate for the host galaxy.
However, we can estimate the redshift using the empirical relationship
$|\Gamma_{\gamma}| = (2.76\pm0.09)(1+z)^{-0.75\pm0.06}$ (Amati
et~al. 2002). For GRB~051117A, $\Gamma_{\gamma}=1.83\pm0.07$ (15.0-150~keV),
we estimate a redshift of z$=0.73^{+0.17}_{-0.21}$. This redshift is
consistent with the upper limit placed on z from the strong detection of this
GRB in all of the UVOT filters ($z<1.4$, adopting $\lambda$2200\AA\ as the
central wavelength for UVW2). For a BAT fluence of
$4.6\times10^{-6}$~erg~cm$^{-2}$ we estimate an isotropic gamma-ray energy of
$E_{\rm iso}=7.8\times10^{51}$~erg (assuming a WMAP Cosmology of
$H_{0}=$70~km~s$^{-1}$~Mpc$^{-1}$, $\Omega_{\lambda}=0.73$,
$\Omega_{m}=1-\Omega_{\lambda}$). Re-writing the Amati relation (Amati
et~al. 2002) in terms of the isotropic energy, $E_{\rm iso}$, and peak energy,
$E_{\rm peak}$, i.e. $E_{\rm peak}=95(E_{\rm iso}/10^{52} erg)^{0.52}$
(Friedman and Bloom 2005), we derive a rest-frame peak energy of $E_{\rm peak}
> 306$~keV. The total energy radiated in gamma-rays can be estimated from the
Ghirlanda relation (Ghirlanda, Ghissellini and Lazzati, 2004) which for
GRB~051117A gives $E_{\gamma}=7.5\times10^{49}$~erg, which implies a beaming
fraction $f_{\rm b} =0.01$.  Both $E_{iso}$ and $E_{\gamma}$ place GRB~051117A
amongst the low-end of the $E_{iso}$ and $E_{\gamma}$ distributions given in
Frail et~al. (2001).

The predicted jet opening angle and break-timescale for this source are
$0.14^{c}$ and 18.4~days respectively (assumes $\eta = 0.2$, and $n =
0.1$~cm$^{-3}$). This jet-break timescale is far later than the break in the
late-time XRT light-curve and is only just covered by the late time data.

\section{Summary and Conclusions}

GRB051117A is one of the brightest GRBs (in terms of detected counts) observed
through the Narrow Field X-ray Telescope at early times.  The unprecedented
S/N of the early X-ray data has revealed complex temporal behaviour, both in
the light-curve and the spectrum. The lightcurve displays multiple episodic
flaring, indicative of a stochastic process. These flaring episodes result in
an abrupt hardening of the X-ray spectrum which then slowly softens as the
flares decline in intensity. Consequently, we find a significant correlation
between source intensity and those parameters which characterise the
properties of the spectrum, e.g. photon index $\Gamma$, hardness ratio $HR$
and break energy $E_{\rm break}$.  We find no evidence for variation in the
hydrogen column in any of our data.  Since spectral evolution is a
characteristic common to many previously observed GRBs (particularly in the
prompt emission), we suggest that the simplest explanation for the observed
spectro-temporal variations is a model in which the break energy correlates
with source intensity. That is, the break-energy moves to higher energies at
the onset of a flare before declining more slowly as the flare fades.

Finally, given the superb S/N of our data we have searched for the presence of
emission-line features in the early X-ray spectrum. We can rule out the
presence of lines with EW$<$15 eV, assuming an intrinsic line width of 100~eV,
at greater than 3$\sigma$ at the peak of the effective area.


\begin{acknowledgements}

We would like to thank the anonymous referee for very helpful comments which
have helped to improve the work presented here.  This work is supported at the
University of Leicester by the Particle Physics and Astronomy Research Council
(PPARC), at Penn State by NASA contract NAS5-00136, and in Italy by funding
from ASI on contract number I/R/039/04.

\end{acknowledgements}

\email{mrg@star.le.ac.uk}.


\clearpage

\begin{table*}
\begin{center}
\caption{Swift XRT observing log.}
\begin{tabular}{cccr}
\hline\hline
\multicolumn{1}{c}{Obsid} & \multicolumn{1}{c}{Tstart} &
\multicolumn{1}{c}{Tstop} &  \multicolumn{1}{c}{Texp} \\
 & MET (s) & MET (s) & (s)\\
 &         &         &    \\
00164268000 & 153917592.991 & 153924985.900 &  4092.9$^{\dagger}$ \\
00164268001 & 153934003.854 & 153974578.652 & 14893.0 \\
00164268002 & 153974610.738 & 154051198.756 & 30341.0 \\
00020022002 & 154055821.405 & 154305538.688 & 57527.0 \\
00020022004 & 154310791.615 & 154392358.971 & 21640.0 \\
00020023002 & 154397100.274 & 154479238.316 & 22902.0 \\
00020023003 & 154483724.965 & 154739637.492 & 64452.0 \\
00020023004 & 154743975.915 & 154912439.176 & 20419.0 \\
00020023005 & 155004275.072 & 155086497.067 & 18321.0 \\
00020023006 & 155091100.675 & 155173374.540 & 16275.0 \\
00020023007 & 155177778.735 & 155259236.189 & 21247.0 \\
00020023008 & 155264657.803 & 155519998.411 & 70525.0 \\
00020023009 & 155520621.431 & 155589836.745 &  7767.5 \\
00020023010 & 155693247.237 & 155775479.364 & 19183.0 \\
00020023011 & 155780139.059 & 155862711.964 & 19581.0  \\
00020023012 & 155866953.038 & 155971765.193 & 20783.0   \\
\hline
\end{tabular}
\end{center}
\noindent $\dagger$ includes both WT and PC mode data.
\end{table*}

\begin{table*}
\begin{center}
\caption{Parameterisation of the early X-ray
light-curve. The underlying powerlaw has a slope of $\alpha = 0.77^{+0.08}_{-0.06}$\label{table_flares}}
\begin{tabular}{rrrr}
\hline\hline
flare no. & \multicolumn{1}{c}{T$_{\rm start}$}&
\multicolumn{1}{c}{T$_{\rm peak}$}&
 \multicolumn{1}{c}{e-folding timescale} \\
& \multicolumn{1}{c}{(s)} & \multicolumn{1}{c}{(s)} & \multicolumn{1}{c}{(s)} \\
& & & \\
a & $0$     &   $144^{+6}_{-5}$   &   $80^{+3}_{-2}$  \\
b & $298^{+8}_{-7}$    &   $351^{+10}_{-2}$   &   $47^{+28}_{-29}$  \\
c & $243^{+32}_{-40}$   &   $331^{+2}_{-6}$ &$0.03^{0.01}_{-0.02}$  \\
d & $417^{+3}_{-18}$  &   $492^{+14}_{-7}$   &   $101^{+14}_{-7}$  \\
e & $346^{+10}_{-8}$    &   $441^{+3}_{-6}$   &   $18^{+10}_{-10}$  \\
f & $572^{+8}_{-8}$   &   $606^{+5}_{-9}$   &  $68^{+22}_{-20}$  \\
g & $810^{+7}_{-8}$    &   $969^{+9}_{-9}$   &  $53^{+7}_{-7}$ \\
h & $978^{+33}_{-27}$  &  $1112^{+6}_{-4}$   &   $58^{+6}_{-8}$  \\
i & $1469^{+14}_{-16}$ &  $1559^{+20}_{-14}$ &  $171^{+71}_{-56}$  \\
j & $1255^{+3}_{-2}$   &  $1325^{+3}_{-3}$   &  $200^{+15}_{-14}$ \\
& & & \\
\hline
\end{tabular}
\end{center}
\end{table*}


\begin{table*}
\begin{center}
\caption{Swift XRT spectral fits. Quoted errors are 90\% confidence on 1
  interesting parameter.}
\begin{tabular}{ccccc}
\hline\hline
 & \multicolumn{4}{c}{GRB~0501117A} \\
& \multicolumn{4}{c}{Model 1: $wa*(wa*po^{\dagger}$) - Nh$_{\rm
 Gal}=1.82\times10^{20}$~cm$^{-2}$, Nh$_{\rm excess} = 2.1\times 10^{21}$~cm$^{-2}$, $\Gamma$ untied} \\
& & & & \\
Mode             & WT orb 1 & WT orb 2 & PC orb 2 & PC orb 3 onward  \\
$\Gamma$         & $2.23\pm0.02$ & 2.39$^{+0.16}_{-0.15}$ & 2.45$^{+0.08}_{-0.07}$ & 2.52$\pm^{+0.15}_{-0.08}$ \\
$\chi^{2}_{\nu}$/dof & 666/676 & & & \\
& & & & \\
\hline
& \multicolumn{4}{c}{Model 2: $wa*(wa*po$) - $\Gamma=2.24$, Nh$_{\rm
 Gal}=1.82\times10^{20}$~cm$^{-2}$, Nh$_{\rm excess}$ untied} \\
& & & & \\ 
Mode             & WT orb 1 & WT orb 2 & PC orb 2 & PC orb 3 onward        \\
$N_{\rm excess}$     &  0.21  & 0.17 & 0.12 & 0.11 \\
$\chi^{2}_{\nu}$/dof & 664/676 & & & \\
& & & & \\
\hline
& \multicolumn{4}{c}{Model 3: $wa*(wa*bknpow)$ - $\Gamma-pre=1.53\pm0.2$,
 $\Gamma-post=2.17\pm0.03$ } \\
& \multicolumn{4}{c}{ and tied,  Nh$_{\rm
 Gal}=1.82\times10^{20}$~cm$^{-2}$, Nh$_{\rm excess}=1.32\pm0.02\times10^{21}$~cm$^{-2}$} \\
& & & & \\
Mode              & WT orb 1 & WT orb 2 & PC orb 2 & PC orb 3 onward \\
E$_{break}$       & 1.10$\pm0.06$ & 0.95$^{+0.30}_{-\infty}$ &
 0.56$^{+0.26}_{-\infty}$ & $0.60^{+0.08}_{-\infty}$ \\
$\chi^{2}_{\nu}$  & 669/674 & & & \\
\hline
& \multicolumn{4}{c}{Model 4: $wa*(wa*bknpow)$ } \\
& \multicolumn{4}{c}{ $\Gamma$ untied,  Nh$_{\rm
 Gal}=1.82\times10^{20}$~cm$^{-2}$, Nh$_{\rm excess}=1.32\pm0.02\times10^{21}$~cm$^{-2}$} \\
& & & & \\
Mode              & WT orb 1 & WT orb 2 & PC orb 2 & PC orb 3 onward \\
E$_{break}$       & 1.10$\pm0.06$ & 0.95$^{+0.30}_{-\infty}$ &
 0.56$^{+0.26}_{-\infty}$ & $0.60^{+0.08}_{-\infty}$ \\
$\chi^{2}_{\nu}$  & 669/674 & & & \\
& & & & \\
\hline
\end{tabular}
\end{center}
\noindent $\dagger$ Here we use the standard XSPEC notation where $wa*po$
means absorbed powerlaw.\newline  
\end{table*}

\begin{table*}
\scriptsize
\begin{center}
\caption{UVOT multicolour photometry}
\begin{tabular}{rrcrrr} 
\hline \hline
$T_\mathrm{mid}$ & Exposure & Filter &
 Mag & Err & Significance \\
(s) & (s) &  &
  &  &  \\
& & & & & \\
      503 &        50 &   $B$ & $>$20.5  & -- &   2.0 \\
      882 &        50 &   $B$ &    20.47 &    0.56 &   2.0 \\
     1487 &       100 &   $B$ &    20.06 &    0.34 &   3.4 \\
      341 &        50 &  UVM2 &    19.26 &    0.52 &   2.2 \\
      720 &        50 &  UVM2 &    19.43 &    0.57 &   2.1 \\
     1174 &       100 &  UVM2 &    20.01 &    0.61 &   2.3 \\
     5227 &       841 &  UVM2 &    20.81 &    0.30 &   3.7 \\
      448 &        50 &   $U$ &    19.08 &    0.33 &   3.6 \\
      828 &        50 &   $U$ &    18.89 &    0.29 &   4.0 \\
     1382 &       100 &   $U$ &    19.11 &    0.25 &   4.6 \\
     7178 &       686 &   $U$ &    20.50 &    0.25 &   4.4 \\
      211 &       200 &   $V$ &    19.66 &    0.33 &   3.4 \\
     1188 &       239 &   $V$ &    20.19 &    0.50 &   2.2 \\
      395 &        50 &  UVW1 & $>$19.7  & -- &   2.0 \\
      774 &        50 &  UVW1 &    19.27 &    0.46 &   2.5 \\
     1279 &       100 &  UVW1 &    19.70 &    0.41 &   2.7 \\
     6378 &       900 &  UVW1 &    20.18 &    0.17 &   6.4 \\
     1115 &       199 &  UVW2 &    20.36 &    0.52 &   2.3 \\
      557 &        50 & White &    19.25 &    0.22 &   5.5 \\
      936 &        50 & White &    19.26 &    0.22 &   5.6 \\
    23591 &      7673 &   $V$ &    22.47 &    0.52 &   2.2 \\
    46705 &      7694 &   $V$ &    22.24 &    0.42 &   2.6 \\
    69961 &      6523 &   $V$ &    21.85 &    0.31 &   3.6 \\
    87410 &      6307 &   $V$ &    22.23 &    0.45 &   2.5 \\
   104641 &      6872 &   $V$ &    22.28 &    0.47 &   2.4 \\
   124381 &      8197 &   $V$ & $>$22.6  & -- &   2.0 \\
   228052 &      4957 & White &    21.05 &    0.10 &  12.0 \\
   242588 &      4995 & White &    21.17 &    0.10 &  11.7 \\
   251297 &      7225 & White &    21.19 &    0.10 &  13.6 \\
   260713 &      5277 & White &    21.07 &    0.10 &  13.2 \\
   274342 &      4901 & White &    21.46 &    0.14 &   8.7 \\
   283480 &      5916 & White &    21.29 &    0.11 &  11.3 \\
   294677 &      7159 & White &    21.53 &    0.11 &  10.4 \\
   304093 &      3817 & White &    21.41 &    0.15 &   7.9 \\
   326447 &      5026 & White &    21.89 &    0.20 &   5.8 \\
   335930 &      7276 & White &    21.72 &    0.14 &   8.7 \\
   350046 &      6887 & White &    21.89 &    0.17 &   7.1 \\
   434098 &     20598 & White &    22.49 &    0.17 &   6.9 \\
   520639 &     21926 & White &    22.33 &    0.14 &   8.4 \\
\\
\hline
\end{tabular}
\end{center}
\end{table*}




\begin{thebibliography}{99}

\bibitem{000}Abbey, A.F., Carpenter, J., Read, A., et~al. 2005, Proceedings of
  ``The X-ray Universe'' Conference, El Escorial, Spain, 2005, ESA SP-604, Vol
  1, p943.
\bibitem{005}Amati, L., Frontera, F., Tavani, M. et~al. 2002, A\&A 390, 81.
\bibitem{010}Band, D., Matteson, J., Schaefer, B. et~al. 1993 ApJ 413, 281.
\bibitem{015}Band, D., Barthelmy, S., Beardmore A. et~al. 2005, GCN 4280.
\bibitem{020}Barthelmy, S.D. 2004 Proc. SPIE Vol 5165, 175.
\bibitem{025}Barthelmy, S.D., Barbier, L.M., Cummings, J.R. et~al. 2005
  Sp. Sc. Rev. 120. 143.
\bibitem{030}Barthelmy, S.D., Chincarini, G. Burrows, D.N. et~al. 2005, Nature
  438, 994.
\bibitem{035}Barthelmy, S.D., Canizzo, J.K. Gehrels, N. et~al. 2005, ApJL 635, L133.
\bibitem{040}Bhat, P.N., Fishman, G.J., Meegan, C.A. et al. 1994, ApJ 426, 604.
\bibitem{045}Burrows, D.N., Hill, J.E., Nousek, J.A. et~al. 2004. SPIE 5165 201
\bibitem{050}Burrows D.N., Hill, J.E., Nousek, J.A. et~al. 2005, Sp. Sc. Rev. 120, 165.
\bibitem{055}Burrows D.N., Romano, P., Godet, O. et~al. 2005b, Proceedings of ``The X-ray Universe''
  Conference, El Escorial, Spain, 2005, ESA SP-604, Vol
  1, p877.
\bibitem{060}Butler, N.R. 2007, ApJ 656, 1001.
\bibitem{065}Butler, N.R., Marshall, H.L.  Ricker, G.R. et al. 2003, ApJ 597,
  1010.
\bibitem{067}Campana, S., Romano, P., Covino, S., Lazzati, D., et~al. 2006 A\&A 449, 61. 
\bibitem{070}Campana, S., Mangano, V., Blustin. A.J. et~al. 2006, Nature 442,
  1008-1010 
\bibitem{075}Crider, A. Liang, E.P., Smith, I.A. et al. 1997, ApJL 479,L39.
\bibitem{080}Cusumano, G., Mangano, V., Chincarini, G. et~al 2006, Nature 440, 164.
\bibitem{085}Cummings, J., Angelini, L., Barthelmy, S. et~al. 2005, GCN 3910.
\bibitem{085}De Luca, A., Melandri, A. Caraveo, P.A. et~al. 2005, A\&A 440, 85.
\bibitem{085}Dermer, C. 2004, ApJ 614, 284.
\bibitem{090}Drenkhahn, G. and Spruit, H.C. 2002, A\&A 391, 1141.
\bibitem{095}Falcone, A.D., Burrows, D.N. Lazzati, D. et~al. 2006, ApJ 641, 1010.
\bibitem{100}Fan, Y.Z. and Wei, D.M. 2005, MNRAS 364, L42.
\bibitem{105}Ford, L.A., Band, D.L., Matteson, J.L. et al. 1995, ApJ 439, 307.
\bibitem{110}Friedman, A.S. and Bloom, J.S. 2005, ApJ 627, 1.
\bibitem{115}Frail, D.A., Kulkarni, S.R., Sari, R. et ~al. 2001, ApJ 562, L55.
\bibitem{117}Fruchter, A.S., Thorsett, S.E., Metzger, M.R. et~al. 1999, ApJ,
  519, L13.
\bibitem{119}Galama, T.J. and Wijers, R.A.M.J. 2001,  ApJL 549, L209.
\bibitem{120}Gehrels, N. Sarazin, C.L., O'Brien, P.T. et~al. 2005, Nature 437, 851.
\bibitem{125}Gehrels, N., Chincarini, G. Giommi, P. et~al. 2004, ApJ 611, 1005.
\bibitem{127}Gendre, B., Corsi, A., Piro, L. 2006, A\&A 455, 803. 
\bibitem{130}Ghirlanda, G. Ghisellini, G. and Lazzati D. 2004, ApJ 616, 331.
\bibitem{133}Goad, M.R., Osborne, J.P., Beardmore, A.P., Godet, O., Page,
  K. 2006, In: Gamma-Ray Bursts in the Swift Era, S.S. Holt, N. Gehrels,
  J.A. Nousek, Eds. 2006 American Institute of Physics.
\bibitem{135}Goad, M.R., Tagliaferri, G., Page, K.L. et~al. 2006, A\&A 449, 89.
\bibitem{140}Goad, M.R., Page, K., Burrows, D., Hurley, K., Chester, M. 2005b, GCN 4287.
\bibitem{145}Gao, W.H., and Wei, D.M. 2005, Chin J. Astron Astrophys Vol 5, 6,
  571.
\bibitem{147}Guetta, D., D'Elia, V., Fiore, F., Conciatore, M.L., Antonelli,
  A., and Stella, L. 2006, In : Il Nuovo Cimeto.
\bibitem{150}Hurkett, C. et~al. 2007, MNRAS, in press.
\bibitem{170}King, A., O'Brien, P.T., Goad, M.R., Osborne, J.P., Olsson,E. and 
Page, K.L. 2005, ApJ 630, L113.
\bibitem{155}Kobayashi, S., Piran, T. and Sari, R. 1997, ApJ 490, 92.
\bibitem{175}Krimm, H.A., Barthelmy, S.D., Chester, M.M. et~al. 2007 GCN 6005.
\bibitem{160}Kumar, P., McMahon, E., Panaitescu, A. et~al. 2007, MNRAS 376, L57.
\bibitem{160}Kumar, P. and Narayan, R. 2003, ApJ 584, 895.
\bibitem{165}Kumar, P. and Panaitescu, A., 2000, ApJ 541, L51.

\bibitem{175}Holland, S.T. 2005, GCN 4301.
\bibitem{175}Holland, S.T., Barthelmy, S., Burrows, D.N. et~al. 2006, GCN 4570.
\bibitem{180}Lamb, D., Donaghy, T. Q., Graziani, C. 2005, ApJ 620, 355.
\bibitem{180}Lazzati, D., and Perna, R. 2007, MNRAS 375, L46.
\bibitem{185}Lazzati D., and Perna, R. 2002, MNRAS 330, 383.
\bibitem{185}Lazzati D., Ramirez-Ruiz E., Rees, M.J. 2002 ApJ 572, L57
\bibitem{190}Lehto, H.J. 1989, In {Symposium on Two Topics in X-ray Astronomy,
  Bolgna, Italy, Sep 12-13 1989 (ESA SP-296 NOV 1989)}.
\bibitem{200}M{\'e}sz{\'a}ros, P. \& Rees, M.J. 1997, ApJ 476, 232.
\bibitem{205}Moretti, A., Perri, M., Capalbi, M.  et~al. 2006, A\&A 448, L9.
\bibitem{210}Nakar, E. and Granot, J. 2006, astro-ph/0606011.
\bibitem{215}Nousek, J. A., Kouveliotou, C., Grupe, D. et~al. 2006, 642, 389.
\bibitem{220}Pei, Y.C. 1992, ApJ 395, 130.
\bibitem{225}Page, K. L., Barthelmy, S.D., Beardmore, A.P. et~al. 2006, GCN 5823.
\bibitem{225}Piro, L., Costa, E., Feroci, M.  et al. 1999 ApJ 514, L73
\bibitem{230}Piro, L., Garmire, G., Garcia, M. et al. 2000, Science, 290, 955.
\bibitem{231}Prochaska, J.X., Bloom, J.S., Chen, H.-W. et~al. 2004, ApJ 611, 200.
\bibitem{235}Rees, M. and M{\'e}sz{\'a}ros, P. 1994, ApJL 430, L93. 
\bibitem{240}Rees, M. and M{\'e}sz{\'a}ros, P. 2000, ApJL 545, L73. 
\bibitem{245}Rees, M.J., and M{\'e}sz{\'a}ros, P., 2005, ApJ 628, 847.
\bibitem{250}Reeves, J.N., Watson, D., Osborne, J.P. et al. 2002, Nature 416, 512.
\bibitem{255}Ryde, F. and Petrosian, V. 2002, ApJ 579, 290.
\bibitem{260}Roming, P.W.A., Kennedy, T.E., Mason, K.O.  et~al. 2005, SSR 120, 95.
\bibitem{262}Stratta, G. Fiore, F., Antonelli, L.A. Piro, L. and De Pasquale,
  M. 2004, aPj 608, 846.
\bibitem{265}Tagliaferri, G., Goad, M., Chincarini, G. et~al. 2005, Nature, 436, 985. astro-ph/0506355
\bibitem{270}Watson, D., Reeves, J.N. Hjorth, J. et al. 2003, ApJ 595, L29.
\bibitem{275}Willingale, R., O'Brien, P.T., Osborne, J.P., Godet, O., Page, K.L. et~al. 2007, ApJ in press.
\bibitem{280}Woosley, S.E. and Bloom, J.S. 2006 ARA\&A, 44, 507.
\bibitem{285}Yoshida, A., Namiki, M. Yonetoku, D. et~al. 2001, ApJ  557, L27.
\bibitem{290}Zhang, B. and M{\'e}sz{\'a}ros, P. 2004 IJMP 19, 2385.


\end{thebibliography}
\end{document}